\documentclass[aps,pre,nofootinbib,twocolumn,superscriptaddress,
               amsmath,amssymb,floatfix]{revtex4}

% ...

\usepackage{amsmath} \usepackage{epsfig} \usepackage{bm}

\usepackage{graphicx}

\usepackage{mathrsfs}                                                 
\usepackage{color}
\usepackage{epsfig}
\usepackage{amssymb}

\newcommand{\dd}{\text{d}}

\newcommand{\ee}{\text{e}}

\newcommand{\bx}{\text{\bf x}}
\newcommand{\br}{\text{\bf r}}
\newcommand{\by}{\text{\bf y}}
\newcommand{\bz}{\text{\bf z}}

\newcommand{\bnabla}{\boldsymbol{\nabla}}

\begin{document}

\title{Dynamic transition in an atomic glass former: a molecular dynamics evidence}

\author{Estelle Pitard}

\affiliation{Laboratoire des Collo\"{\i}des, Verres et NanoMat\'eriaux (CNRS UMR 5587),
 Universit\'e de Montpellier II, place Eug\`ene Bataillon, 34095 Montpellier cedex 5, France}

\author{Vivien Lecomte}

\affiliation{Laboratoire de Probabilit\'es et Mod\`eles Al\'eatoires
  (CNRS UMR 7599), Universit\'e Paris Diderot - Paris 7 et Universit\'e
  Pierre et Marie Curie - Paris 6, Site Chevaleret, CC 7012, 75205
  Paris cedex 13, France}

\author{Fr\'ed\'eric van Wijland}

\affiliation{Laboratoire Mati\`ere et Syst\`emes Complexes (CNRS UMR
7057), Universit\'e Paris Diderot - Paris 7, 10 rue Alice Domon et L\'eonie Duquet, 75205 Paris cedex
13, France}

\begin{abstract}
We find that a Lennard-Jones mixture displays a
 dynamic phase transition between an active regime and an inactive one.
  By means of molecular dynamics simulations  
  and of a finite-size study,
   we show that the space time dynamics in the supercooled regime 
coincides with   a dynamic 
   first order transition point. 
\end{abstract}

\maketitle
In order to realize that a physical system has fallen 
into a glassy state one must either drive the system out of equilibrium, 
or investigate its relaxation properties. It is indeed a well-known fact
 that  since no static signature is available, one has to resort to
  nonequilibrium protocols or measurements to identify a glassy state.
   The supercooled regime, which sets in before the material actually
    freezes into a solid glass, is also characterized by anomalous temporal properties,
     such as the increase in the viscosity, and ageing phenomena during relaxation processes.
      The density-density autocorrelation function at a microscopic scale,
       instead of exhibiting a straightforward exponential relaxation,
develops a plateau (over a duration conventionally called $\tau_\alpha$) before relaxing  to zero.
 The present work is concerned with the dynamical properties in the  supercooled regime.
  The idea that dynamical heterogeneities, long-lived large scale spatial structures,
   are the trademark of the slow and intermittent dynamics in the supercooled regime dates back to
  almost thirty years \cite{FA}
  and many developments have occurred since \cite{hetero,garrahanchandler,merollegarrahanchandler}.
Here, instead of resorting to local probes such as the van Hove function, 
the self-intermediate scattering function, or the nonlinear dynamical susceptibility, 
- {\it i.e} on multi-point correlation functions in space and time-, 
we prefer to rely on a global characterization of the system's dynamics.

More recently, it was advocated by Garrahan
 {\it et al.}\cite{garrahanjacklecomtepitardvanduijvendijkvanwijland}
  that, at the level of Kinetically Constrained Models (KCM),
   dynamical heterogeneities appeared as the consequence of a universal phase transition mechanism,
largely independent of the specific model under consideration.
 The phase transitions at work are not of a conventional type:
  they occur in the space of 
  trajectories of realizations the system has
   followed over a large time duration, instead of occurring
    in the space of available configurations,
         as the liquid-gas transition, or a putative thermodynamic glass transition, for instance, do.
There exists a well-established body of mathematical physics
 literature  to analyze these phase transitions,
  which is based, in spirit, on the thermodynamic formalism of Ruelle~\cite{ruelle,gaspard},
   and which was adapted~\cite{appertlecomtevanwijland-2} and
    then exploited~\cite{garrahanjacklecomtepitardvanduijvendijkvanwijland}
in numerical and analytical studies of the KCM's.
 The idea behind these studies is to follow the statistics,
  upon sampling the physical realizations of the system over a large time duration,
   of a space and time extensive observable. This observable
    characterizes the level of dynamical activity in
     the course of the system's time evolution.
At this stage, the term activity
 refers to the intuitive picture
  that an active trajectory is
  characterized by
  many  collective rearrangements needed to escape from energy basins,
     whereas an inactive trajectory is
    dominated by the rapid in-cage rattling motion of the particles without collective rearrangements.
This activity observable is used to partition trajectories into two groups
 --the inactive and the active ones-- depending on whether their activity is above or below the average activity.

An important step forward in probing the relevance
 of the dynamic phase transition scenario to realistic glasses
  is the work of Hedges {\it et al.}~\cite{hedgesjackgarrahanchandler}
   in which an atomistic model was considered.
    The system studied by these authors is a
     mixture of two species of particles interacting via a Lennard-Jones potential
       otherwise known as the Kob-Andersen (KA) model \cite{kobandersen},
        and which has been shown
	 to fall easily into a glassy state without crystallizing.
Endowed with a Monte-Carlo or a Newtonian dynamics,
 Hedges {\it et al.} implemented the Transition Path Sampling (TPS)
  method to produce the probability distribution function (pdf)
  of the activity, at finite times.
   These authors succeeded in showing that for a finite system,
    as the observation time was increased, the pdf of the activity
     develops a bimodal structure.
In the light of previous works on lattice models for glass formers,
 they interpreted their result as an evidence for a
  {\it bona fide} dynamic phase transition,
   expected to occur once the infinite size
    and infinite observation time limits are reached.
While the work of Hedges {\it et al.}
 is a definite breakthrough towards a better
  understanding of atomistic glass formers,
   it leaves a number of questions unanswered.
These are the subject of the present work.
 Dynamic phase transitions occur in the large time limit,
  as exemplified by the pioneering works of Ott {\it et al.}~\cite{ottwhithersyorke},
   Cvitanovi\'c~\cite{artusocvitanovickenny} or Sz\'epfalusy and T\'el~\cite{szepfalusytel}
    (see Beck and Schl\"ogl~\cite{beckschlogl} for further references).
     The dynamic transitions at work in glass formers are of a subtler form,
      since they only emerge upon, in addition, considering the infinite system size limit.\\

Our first goal in this letter is to show the existence of a phase transition.
Our second goal is to be more precise about
 the location of the phase transition itself. This is indeed  a central issue:
  if the critical point is away from the typical measurable activity then
   the transition scenario is at best a crossover.
    If, on the contrary, it is shown that
     the phase transition actually takes
      place for typical trajectories then experimentally observable
       consequences are expected.
Our third goal is to provide a rigorous definition of the activity. 
 It is our feeling that deeper insight will
  be gained if the notion of activity can be
   characterized by standard physical concepts
    (like forces between particles)
     rather than on phenomenological considerations.

\begin{figure}[htbp]
\begin{center}
\includegraphics[width=.4\columnwidth]{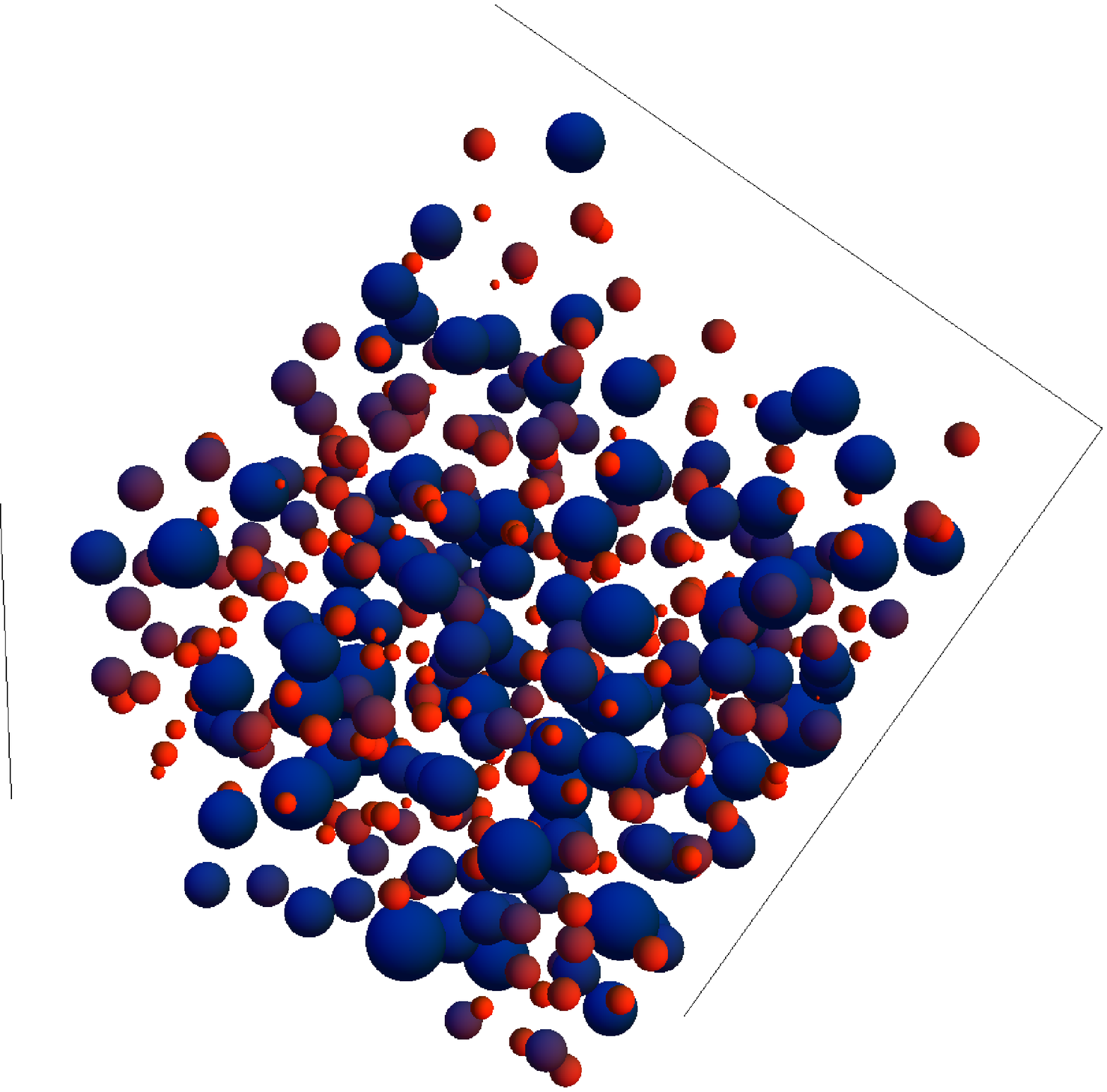}
\quad
\quad
\includegraphics[width=.4\columnwidth]{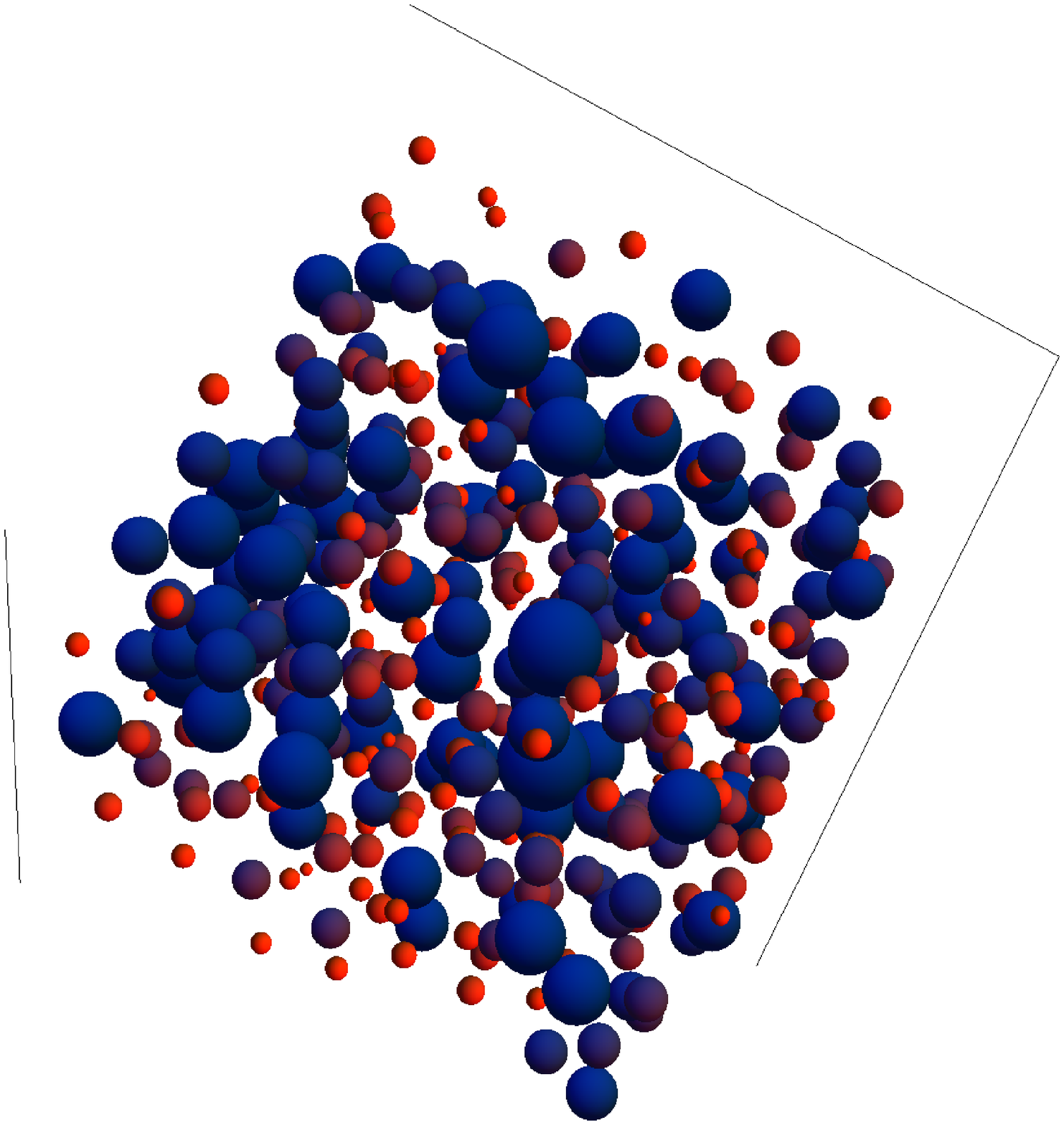}
\vspace*{-7mm}
\end{center}
\caption{In these snapshots, the diameter of the particles quantifies the local activity: small points refer to mobile active
particles, whereas large blobs refer to blocked inactive particles. Red (blue) indicates that the activity is larger (smaller) than the median one.
Left: Activity map in a typical configuration from a trajectory with excess activity
 with respect  to the average activity; there is a large number of mobile particles though blocked particles also exist.
   Right: Activity map in a typical configuration from a trajectory with less activity
 than the average activity, large blobs dominate.
   \label{activitysnapshot}}
\end{figure}

In the present work, we consider a KA mixture,
  which we study by extending the methods of Molecular Dynamics (MD)
   to the study of temporal large deviations. By implementing a MD
    version of the cloning algorithm of Giardin\`a, Kurchan and Peliti~\cite{giardinakurchanpeliti}
     we are able to follow the statistics of the large deviations of the activity.
 Our activity observable
  is a physically transparent quantity
   which can be related to the rate at
    which the system escapes from a given
     location in phase space.
As we shall soon see,
 the typical trajectories
  of the system in phase space
   are characterized by strong space time heterogeneities
    --the so-called dynamical heterogeneities-- 
    which appear to be the by-product of an underlying
     first order dynamical transition in which
      the activity plays the role of an order parameter.
       A typical realization lies at a first-order transition point,
        which is thus characterized by the coexistence of competing families of trajectories.
Active trajectories, in which a time and space extensive number of events where a particle
 escapes from its local cage, coexist with inactive trajectories
  in which localization of the particles within their local
   neighborhood dominates the dynamics. Without going further
    into the mathematical details of what the activity is,
     we present in figure \ref{activitysnapshot}
      a snapshot of a the activity map in
       a three-dimensional KA mixture in three distinct situations.
       
We now carry out our program and we begin
 by defining an activity observable:
  given a set of particles interacting
   via a two-body potential $V(\br_i-\br_j)$,
    in which each particle $i$ 
    is subjected to a force
     ${\bf F}_i=-\sum_{j\neq i}\bnabla V(\br_i-\br_j)$, we introduce the observable:
      
\begin{equation}\label{defVeff}
V_\text{eff}=\sum_i\left[\frac{\beta}{4}{\bf F}_i^2+\frac 12 \bnabla_{\br_i}\cdot {\bf F}_i\right]
\end{equation}

The combination
 $\frac{\beta}{4}{\bf F}_i^2+\frac 12 \bnabla_{\br_i}\cdot {\bf F}_i$
  can be interpreted as the activity of a single particle.
   The quantity $V_\text{eff}$ in \eqref{defVeff}
    appears in the study of Brownian interacting particles
    and measures the tendency for dynamical trajectories
    to evolve away from a given configuration.
     Indeed, it can be argued~\cite{autierifacciolisegapederivaorland}
      that the quantity $\exp(-\beta V_\text{eff}\dd t)$
       is 
       proportional to the probability $P(x,t+dt|x,t)$ that the system has
        stayed in its configuration between
	 $t$ and $t+\dd t$.
This means that $\beta |V_\text{eff}|$
 is the rate at which the
  system escapes its configuration.
  To understand how~\eqref{defVeff} is related to this escape rate
  one writes $P(x,t+dt|x,t)=\langle x| e^{- \hat H \dd t}|x \rangle$
  where $\hat H$ is the Fokker-Planck operator of evolution of the
  system. Using the detailed balance property of the dynamics, $\hat
  H$ can be symmetrized and this leads to
  $P(x,t+dt|x,t)\sim\exp(-\beta V_\text{eff}\dd t)$, using
  standard path-integral representation of the propagator $\langle x| e^{- \hat H \dd t}|x \rangle$
  (see~\cite{autierifacciolisegapederivaorland} for details).
   Besides, if one regularizes
    the Brownian dynamics in the form of hops on a lattice,
     our activity can be
      viewed as the continuum analog of the activity introduced in Lecomte
       {\it et al.}~\cite{lecomteappertvanwijland-1}
        which counted the number of
	 configuration changes
	  undergone by a system over a given time interval.
Perhaps more interestingly still,
 $V_\text{eff}$ also appears as
  the continuum limit analog of
   the dynamical complexity
    (a trajectory-dependent Kolmogorov-Sinai entropy~\cite{appertlecomtevanwijland-2})
     which further clarifies its conceptual value.
      However, hand-waving arguments lead to
       a more concrete understanding of the value of the effective potential $V_\text{eff}$.      
Indeed, in the expression \eqref{defVeff}
  minimizing the first term in the right hand side
   drives the system away from
   regions of phase space where forces are nonzero.
    In other words, it tends to favor
     mechanical equilibrium configurations
      (whether stable or metastable).
The second term in the right hand side of \eqref{defVeff},
can be positive or negative: if negative,
  it will favor local minima, all the more so as they are deep and steep; if positive, it will select local maxima.
  As a result, there will be  two classes of trajectories, according to the values of 
$V_\text{eff}$. It can be shown that the average equilibrium value of $V_\text{eff}$
is $\langle V_\text{eff}\rangle=-\beta/4\sum_i F_i^2$, which is always negative. For minimal (negative) values of   $V_\text{eff}$, trajectories will explore deep energy basins, for
maximal values of $V_\text{eff}$, trajectories
will explore local maxima of the energy landscape.

   The total activity is defined as
   $K(t)=\int_0^t \dd t' V_\text{eff}(t')$.
   In terms of the local density $\rho(\bx,t)$, our activity reads:
   
\begin{equation}\begin{split}
K=&\int_0^t\dd t\Big[\frac{1}{4T}\int_{\bx,\by,\bz}\!\!\!\!\!\!\!\!\!\!\!\!\rho(\bx,t)\rho(\by,t)\rho(\bz,t)
\bnabla V(\bx-\by)\bnabla V(\bx-\bz)\\&-\frac{1}{2}\int_{\bx,\by}\!\!\!\!\!\!\rho(\bx,t)\rho(\by,t)\Delta V(\bx-\by)\Big]
\end{split}\end{equation}
It is now apparent that $V_\text{eff}$
 involves three-body effective interactions,
  a fact that we {\it a posteriori}
   interpret by realizing that  in order
to escape from an energy basin via a collective rearrangement,      
 multi-body interaction terms are needed.

\begin{figure}
\includegraphics[width=7cm]{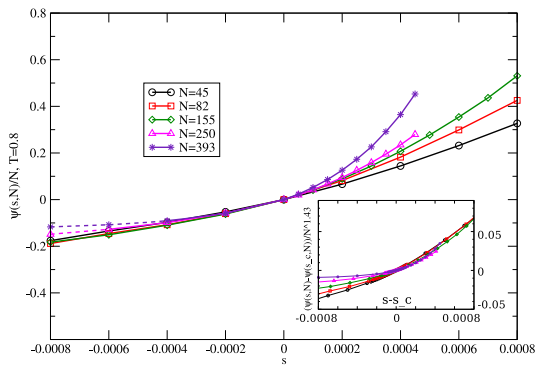}
\caption{The dynamical free energy per particle $\frac{\psi(s,N)}{N}$
  is shown as a function of $s$ for increasing sizes $N=45$, $82$,
  $155$, $250$ and $393$.  While the curves collapse onto a single
  master curve at $s<0$, a strong size-dependent behavior is observed
  at $s>0$.  On the $s>0$ side, the curves display a rapid variation
  at a typical value $s_c(N)$ which decreases as $N$ increases,
  leading to the putative building up of a singularity of
  $\psi(s,N)/N$ at $s=0$ as $N\to\infty$. The rescaled
  $[\psi(s,N)-\psi(s_c,N)]/N^{1+\alpha}$ with $\alpha\simeq 0.43$ is
  shown in the inset: on the inactive side $s>0$ the curves collapse
  onto a single curve, exemplifying the large $N$ finite size scaling.
  Applying the cloning algorithm requires the $| s\, V_\text{eff}\, \dd t |$
  product to be infinitesimally small (in principle), which at the
  left end of the graph and for the largest two values of N is hardly
  verified (dashed lines). For these points, this product reaches $.5$ which accounts
  for the increasing error.  }
\label{psides}
\vspace*{-5mm}
\end{figure}

Given the activity we have just defined,
 we set out to determine the distribution
  of this quantity over a large number of realizations of the process,
   that is, given the dynamics is deterministic,
    by sampling over a large number of initial states drawn
     from the equilibrium Boltzmann distribution.
Sampling the full distribution of the activity requires sufficient statistics
 beyond its typical values. For that reason we have preferred
  to work in an ensemble of trajectories in which the average activity,
   rather than the activity, is fixed. In this canonical version,
    we consider a large number of systems evolving in parallel,
     and between $t$ and $t+\dd t$ we remove or add a system in a given configuration 
     with a cloning factor equal to $\exp(-s  V_\text{eff}(t)\dd t)$.
How to perform a numerical simulation
 in a canonical ensemble of time realizations
  was explained in Giardin\`a {\it et al.}\cite{giardinakurchanpeliti}'s
   cloning algorithm.
    We combined the cloning algorithm with the molecular dynamics simulation
     of the interacting particles.
      Choosing a positive value of $s$ allows one 
      to focus on time realizations with a lower-than-average activity, 
      while a negative $s$ selects over-active trajectories, 
      and $s$ close to zero samples typical activity trajectories.
       To this day,
        no one has been able to endow the parameter conjugate to the activity $s$
	 with a concrete physical meaning, 
	 so that the only physically realizable value of an $s$-ensemble
	  is achieved for $s=0$.
As will shortly become clear,
 analyzing the $s=0$ properties
  gains enormously from being able to vary $s$ away from 0,
   a theorist's privilege.
    The distribution of the activity $K$ 
    is fully encoded in its generating function $Z(s,t,N)=\langle \ee^{-s K}\rangle$,
     or, alternatively, in the corresponding cumulant generating function $\ln Z$.
We define the intensive dynamical free energy $\psi(s,N)=\lim_{t\to\infty}\frac{\ln Z}{t}$
 for any given finite size.
  From the knowledge of the function $\psi(s,N)$
   we can reconstruct the properties of the activity $K$.
    In numerical terms, in view of the cloning algorithm used,
     $\psi$ stands for the growth rate of the population of
      systems evolving in parallel.
       The details of the simulations are as follows. 
We used the A-B mixture of \cite{kobandersen}  for samples of different sizes. The samples all had the approximate ratio 80/20 for A/B
particles. They were first prepared at equilibrium at temperature $T=0.8$ by coupling to a stochastic heat bath; the time step was
$\dd t=2.10^{-2}$. During a very long
simulation at equilibrium, the $N_c$ clones where prepared. Then the cloning algorithm 
of Giardin\`a {\it et al.}\cite{giardinakurchanpeliti} was performed with a small coupling
to the heat bath, necessary to provide some stochasticity to the exploration of trajectory space, as explained in 
\cite{giardinakurchanpeliti,tailleurkurchan} for
deterministic systems. In our simulations, the convergence was checked by varying the number of clones and the duration of
the simulations. For $N=45,82$ the best results were found for $N_c=1000$ and $\tau=10 \tau_{\alpha}$ ($\tau_{\alpha}$
is the relaxation time); for $N=155, 250, 393$ the best results were found for
$N_c=500$ and $\tau=20 \tau_{\alpha}$. 
Fluctuations in the space of clones would have required for the largest system sizes that we substantially increase $N_c$ as the tails of the distribution are sampled (extreme values of $s$).

\begin{figure}
\includegraphics[width=6.0cm]{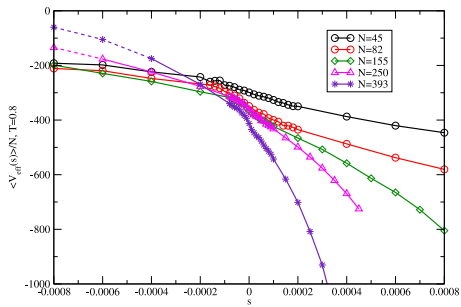}
\vspace*{-3mm}
\caption{The average activity per particle
 $\langle V_\text{eff}\rangle/N$ is measured in an
  ensemble of trajectories biased by a weight
   $\ee^{-s K}$ and is plotted as a function of $s$.
    While the $s<0$ regime shows convergence to a size independent
     limit of an order comparable to its equilibrium value,
      the $s>0$ side confirms a strongly size-dependent behavior as $V_\text{eff}$
       abruptly decreases to negative values.
        This sudden drop in $V_\text{eff}$ indicates that the system freezes into inactive states.}
\label{Veffdes}
\vspace*{-5mm}
\end{figure}

\begin{figure}
\includegraphics[width=6.0cm,angle=270]{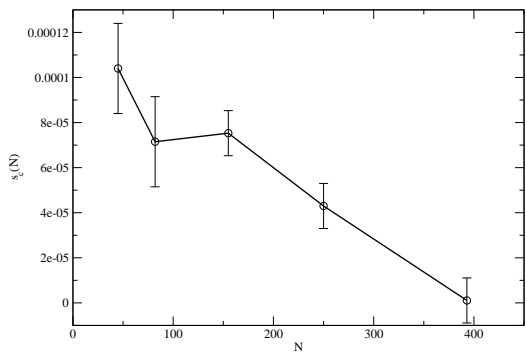}
\vspace*{-4mm}
\caption{$s_c(N)$ 
 as a function of $N$.}
\label{scden}
\vspace*{-4mm}
\end{figure}

We now present the central result of the work, namely the dynamical
free energy $\psi(s,N)$ as a function of $s$ for increasing system
sizes $N$.  Figure \ref{psides} shows that the dynamical free energy
per particle $\frac{\psi(s,N)}{N}$ settles to a well-defined limiting
value for $s<0$ as $N$ increases.  This means that for $s<0$ there
exists a thermodynamic limit.
On the $s<0$ side, where the activity is above its average value at
$s=0$, which we interpret in terms of active trajectories, the
physical properties are not far from those at $s=0$ as predicted from
the Gibbs equilibrium distribution.  However, a dramatic change in
behavior is observed for positive values of $s$.  Not only do the
dynamical free energies exhibit a strong size dependence, but a severe
change in behavior is also observed on that side of the parameter
space.  For $s>0$, the free energy rapidly increases, and the activity
rapidly decreases.  The location $s_c(N)$ at which this rapid increase
sets shows a strong $N$-dependence: $s_c(N)$ seems to decrease to zero
as $N$ increases.  The precise location of $s_c(N)$ has been
determined thanks to a spline method and by maximizing $\psi''(s)$;
this is illustrated in figure \ref{scden}.  One notes that given the
existence of error bars, there is an overall decrease of $s_c(N)$ as
$N$ increases, but one cannot exclude that $s_c(N)$ converges to a
very small positive value.
As in other examples of first order dynamical
transitions~\cite{garrahanjacklecomtepitardvanduijvendijkvanwijland},
we remark that $s_c(N)$ remains positive for all finite $N$. Indeed,
$\psi'(0)=-\langle V_\text{eff} \rangle$ is proportional to $N$, which
means that $\psi(s)$ also scales as $N$ in the vicinity of $s=0$, and
since $s_c(N)$ marks the transition to a regime where $\psi(s)$ scales
differently with $N$, one has $s_c(N)>0$.

The interpretation in terms of active {\it vs.}  inactive trajectories
is easier when plotting the average activity as a function of $s$.
This is done in figure \ref{Veffdes}, where the average activity is
measured independently: we see that in the vicinity of $s_c(N)$, the
behavior of $\langle V_\text{eff}\rangle/N$ changes abruptly from a
smooth range of values at $s<0$ to a strongly negative range for
$s>0$.  In this inactive regime, through rescaling (see the inset on
figure \ref{psides}), we see that the free energy behaves in
$N^{1+\alpha}$ with $\alpha\simeq 0.43$, in contrast to the purely
extensive behavior of the active regime.
A similar difference of scaling exponents in $N$ is observed in KCMs
in two
dimensions~\cite{garrahanjacklecomtepitardvanduijvendijkvanwijland} at
$s>0$: $\psi(s)$ is of order $L$ for the Fredrickson-Anderson model (FA) while of
order $1$ for the triangular lattice gas (TLG). This behavior is related
to the geometry of the remaining active sites in the system:
those divide up along the border of fully inactive domains in the FA model,
while they are isolated in the TLG.
Our finding that $\alpha$ is non-integer indicates that the KA mixture
adopts configurations with non-trivial geometrical features for the
inactive particles in the inactive phase due to the effective
long-range interactions that develop in the $s>s_c$ states. Since
large values of $|V_{\text{eff}}|$ correspond to inactive histories,
it is expected that $\alpha>0$, as observed.
We note that such finite-size behaviour of the activity with
non-trivial exponents is known to occur in lattice
KCMs~\cite{bodineautoninelli}, where the value of those exponents are
directly related to the nature of configurations appearing in the
inactive state (which differ from those of the active state).

     This allows us to claim that
      there exists a phase transition from inactive to active
       states as $s$ is varied. Our closely related 
       goal was
        to identify the location of the transition.
	 We have shown that $s_c(N)$ displays
	  a strong size-dependence 
	  with an overall decrease for
	    the  range of $N$ values explored.
The two different well-defined scalings $\sim N^{1+\alpha}$
(resp. $\sim N$) for $s>0$ (resp. $s<0$) indicate that we have reached
the large $N$ asymptotics on our range of system sizes (see figure \ref{psides}).
We therefore conclude that the most likely value at which the transition
 takes place in an infinite system is either $s_c(\infty)=0$
 or a very small positive value close to $0$. We were not able to perform simulations for larger values of $N$ due to the very
 large computation times needed.

The first order dynamic transition scenario observed in KCM's
 is thus  confirmed in the atomistic model we have studied.
  As we mentioned earlier, the location of the phase transition
   along the $s$-variable axis is an extremely relevant
    issue since only the value $s=0$ is experimentally accessible.
     Note however that identifying a transition point at $s=0$
      is possible only on the condition that $s$ is varied across 0.
       Therefore, finding evidence for the transition occurring at $s=0$
        renders an experimental observation of the transition a credible achievement.
The lesson to be drawn from the present extensive simulation series is that,
 as expected, it is desirable to work on as small systems as possible,
  yet large enough to allow for collective effects in trajectory space to develop.

This work was supported by a French Ministry of Foreign Affaires
Alliance grant.  VL was supported in part by the Swiss NSF under MaNEP
and Division II.  We greatly benefited from discussions with
R.L. Jack. We thank W. Kob for providing the software for the
preparation of the configurations, and L. Berthier for help with
figure 1.

\end{document}